\documentclass[showpacs,twocolumn,pre]{revtex4}
\usepackage{psfrag,epsfig,amsfonts,amssymb,amsmath,graphicx,slashbox}
\usepackage{dcolumn}

\newcommand{\hr}{{\cal H}}
\newcommand{\ord}{{\cal O}}
\newcommand{\tr}{\mbox{Tr}}
\newcommand{\da}{\Delta_{\! A}}
\newcommand{\dda}{\delta \! A}
\newcommand{\sa}{\sigma_{\! A}}
\newcommand{\mic}{\mathrm{mic}}

\begin{document}

\title{Foundation of Statistical Mechanics under experimentally realistic conditions}

\author{Peter Reimann}
\affiliation{Universit\"at Bielefeld, Fakult\"at f\"ur Physik, 33615 Bielefeld, Germany}

\begin{abstract}
We demonstrate the equilibration of isolated macroscopic 
quantum systems, prepared in non-equilibrium 
mixed states with significant population 
of many energy levels, 
and observed by instruments with 
a reasonably bound working range compared
to the resolution limit.
Both properties are fulfilled under many, 
if not all, experimentally realistic 
conditions.
At equilibrium, the predictions and limitations of 
Statistical Mechanics are recovered.
\end{abstract}


\maketitle
Fundamental aspects of equilibrium Statistical Mechanics
(ESM) are intensely reconsidered at present in the 
context of (almost) integrable many-body quantum 
systems \cite{rig08,exp,rig07,cra08},
bringing back to our attention that very basic
issues are still not satisfactorily understood
\cite{lud58,boc59,ergod,per84,jen85,deu91,sre96,tas98,das03}.
As in every theory, we are faced with the three 
sub-problems to realistically model preparation, 
time evolution, and measurement of a given system.
It is well know and will be worked out in detail
below that the question of experimentally
realistic initial conditions and observables
is much more urgent in the ``derivation'' of
ESM than in most other fields \cite{lud58,ergod,per84,jen85,sre96,realobs}.
Regarding time evolution, we take the widely
(yet not unanimously) accepted viewpoint 
that standard Quantum Mechanics
without any additional ``postulate'' or 
``hypothesis'' must do \cite{lldiu,f9}.
The two key questions are then:
In how far does a non-equilibrium seed
evolve to a stationary long-time behavior 
(``equilibration'')?
In how far is this steady state in 
agreement with the corresponding 
ESM ensemble (``thermalization'')?

Since open systems (interacting and entangled 
with an environment) are not directly tractable 
by standard Quantum Mechanics,
the starting point must be a closed (autonomous)
system (microcanonical framework), incorporating
all relevant thermal baths, reservoirs etc. 
\cite{lldiu,f9}.
Accordingly, the system ``lives'' in some 
Hilbert space $\hr$ and is at any time 
instant $t\geq 0$ in a mixed state  
(including pure states as special case)
$\rho(t)=U_t\rho(0) U_t^\dagger$ 
with propagator $U_t:=\exp\{-iHt/\hbar\}$,
seed $\rho(0)$, and time-independent 
Hamiltonian $H$.
Denoting its eigenfunctions and eigenvalues 
by $|n\rangle$ and $E_n$ ($n=0,1,2,...$) 
and the matrix elements 
$\langle m|\rho(t)|n\rangle$ by $\rho_{mn}(t)$
we thus obtain
\begin{equation}
\rho(t)=\sum \rho_{mn}(0)\, e^{i[E_n-E_m]t/\hbar}
\, |m\rangle\langle n| \ ,
\label{1}
\end{equation}
where the sum runs over all $m,n\geq 0$.
As usual, observables are represented by
Hermitean operators $A$ with expectation 
values $\tr\{\rho(t)A\}$ and,
without loss of generality, are assumed
not to depend explicitly on time.

Generically, the ensemble $\rho(t)$ is 
not stationary right from the 
beginning, in particular for an initial
condition $\rho(0)$ out of equilibrium.
But if the right hand side of (\ref{1})
depends on $t$ initially, it cannot 
approach for large $t$ any time-independent 
``equilibrium ensemble'' whatsoever.
In fact, any mixed state $\rho(t)$ returns arbitrarily
``near'' to its seed $\rho(0)$ for certain, 
sufficiently large time-points $t$, 
and similarly for the expectation values 
$\tr\{\rho(t) A\}$, see Appendix D in
Ref. \cite{hob71}.
More specifically, consider any $\rho(0)$
with at least one $\rho_{mn}(0)\not = 0$
and $\omega:=[E_n-E_m]/\hbar\not =0$.
Chosing 
\begin{equation}
A=B+B^\dagger \ , \  B:=|m\rangle\langle n| /\rho_{mn}(0)
\label{1c}
\end{equation}
it follows that 
$\tr\{\rho(t)A\}=2\, \cos(\omega t)$.
It is thus clearly impossible to ``derive''
(since it is not correct) ESM for arbitrary 
initial conditions and observables.

Our first basic assumption 
concerns the quantity \cite{f8}
\begin{equation}
R:=
\big[\, \sum \rho^2_{nn} (0) \, \big]^{1/3} \leq \big[\, \max_{n} \rho_{nn}(0)\, \big]^{1/3} \ .
\label{1a}
\end{equation} 
According to (\ref{1}), the $\rho_{nn}(t)$
represent the ``occupation probabilities'' of 
the energy eigenstates and are independent of 
$t$.
For a system with $f$ degrees of freedom, there
are roughly $10^{\ord(f)}$ energy levels in every 
interval of $1$J beyond the ground state energy $E_0$ 
\cite{lldiu,f4}. 
For a macroscopic system with
$f=\ord(10^{23})$, the levels are thus unimaginably
dense on any decent energy scale and 
even the most careful experimentalist will not
be able to prepare the system
such that the resulting ensemble $\rho (0)$
populates only a few energy eigenstates 
with significant probabilities \cite{lldiu}.
For example, assume that there are
exactly $10^{(10^{23})}$ energy levels
per J.
Even if the system preparation defines the
energy up to an experimental uncertainty 
of $10^{-(10^{22})}$J, there still 
remain $N:=10^{0.9\cdot 10^{23}}$ energy levels 
which may be occupied with significant 
probabilities.
If all of them are populated equally,
we obtain $\rho_{nn}(0)=1/N$
for $N$ indices $n$ and 
$\rho_{nn}(0)=0$ for all other $n$,
yielding $R\leq 10^{-0.3\cdot 10^{23}}$
according to (\ref{1a}).
If not all $N$ levels are populated
equally, but rather any $\rho_{nn} (0)$ 
may assume arbitrary 
values between zero and $10^{(10^{22})}$
times the average population $1/N$,
Eq. (\ref{1a}) still yields $R\leq 10^{-0.26\cdot 10^{23}}$.
Returning to the general case, we can conclude 
that even if the system energy is fixed up to
an extremely small experimental uncertainty
and even if the energy levels are populated 
extremely unequally, we still expect that 
$R$ will be extremely small, typically
\begin{equation}
R=10^{-\ord(f)} \ .
\label{1b}
\end{equation}
Physical reasons for such a broadly 
spread energy level population include:
the time-energy uncertainty relation,
imperfect (as opposed to ideal) measurements 
during the system preparation ($t<0$),
entanglement processes and a time dependence 
of the Hamiltonian, both caused by that part 
of the environment from which the system 
is isolated for $t\geq 0$ but not for $t<0$.

Given $\rho(0)$, let $\hr_+\subset\hr$ be 
the Hilbert space spanned by those basis 
vectors $|n\rangle$ for which $\rho_{nn}(0)\not = 0$,
\begin{equation}
\hr_+ := \mbox{span}\{|n\rangle\, |\, \rho_{nn}(0)>0\} \ .
\label{1d}
\end{equation}
Exploiting Cauchy-Schwarz's inequality \cite{f10}
\begin{equation}
|\rho_{mn}(t)|^2\leq \rho_{mm}(t)\, \rho_{nn}(t)
\label{5a}
\end{equation}
it follows that $\rho_{nm}(t)=0$ whenever
$\rho_{nn}(t)=0$ or $\rho_{mm}(t)=0$, i.e.
``all non-trivial things are 
expected to happen within $\hr_+$''.

Our second basic assumption
is that $A$ represents an experimental 
device with a {\em finite range} of possible 
outcomes of a measurement within $\hr_+$ 
\cite{rei07},
\begin{equation}
\da := 
\max_{\hr_+} \langle\psi|A|\psi\rangle
- \min_{\hr_+} \langle\psi|A|\psi\rangle 
= a_{max} - a_{min} \ ,
\label{2}
\end{equation}
where the maximization and minimization is over all 
normalized vectors $|\psi\rangle\in \hr_+$ and
where $a_{max}$ and $a_{min}$ are the largest and 
smallest eigenvalues of the restriction/projection of
$A$ onto $\hr_+$.
Moreover we require that this working range $\da$
of the device $A$ is limited to experimentally 
reasonable values compared to its resolution limit 
$\dda$, for instance $\da < 10^{1000}\dda$.

In the worst case, $\hr_+ = \hr$.
However, in many cases the populations $\rho_{nn}(0)$ 
may be safely negligible e.g. beyond some finite upper 
energy threshold, yielding a finite-dimensional
$\hr_+$, while $\hr$ is typically infinite 
dimensional.
Hence, (\ref{2}) will be finite even for
operators $A$ with an unbound spectrum on $\hr$.
In any case, our above specified class of 
admissible observables $A$ clearly includes 
any realistic measurement apparatus.

For the example (\ref{1c}) we can infer 
from (\ref{1a}), (\ref{5a}), (\ref{2}) 
that $\da \geq 2/R^{3/2}$.
Hence, the oscillations 
$\tr\{\rho(t)A\}=2\, \cos(\omega t)$
are below the resolution limit
under our two basic assumptions.
These (or similar) assumptions
seem thus indispensable for taming 
the oscillations in (\ref{1}). 

Given any $\rho(t)$, we define the auxiliary 
operator \cite{f11}
\begin{equation}
\rho_{eq}:=\sum \rho_{nn}(0) |n\rangle\langle n| 
\label{3}
\end{equation}
and focus on the mean square deviation
\begin{equation}
\sa^2:=\overline{ [\tr\{\rho (t)A \}-\tr\{\rho_{eq} A\}]^2} \ ,
\label{4}
\end{equation}
where the overbar indicates an average over all times 
$t\geq 0$.
The two trace-terms in (\ref{4}) can be unified into
$\tr\{\tilde \rho(t)A\}$ with 
$\tilde\rho(t):=\rho(t)-\rho_{eq}$.
Introducing $\tilde A:= A-\min_{\hr_+}\langle\psi|A|\psi\rangle$
we can infer from (\ref{2}) that 
\begin{equation}
0\leq \langle \psi|\tilde A|\psi\rangle \leq \da\  
\mbox{for all normalized}\  |\psi\rangle \in\hr_+ \ .
\label{4a}
\end{equation} 
Since $\langle n|\tilde\rho(t)|n\rangle =0$ it follows
that $\tr \{\tilde \rho(t)\} =0$ and that the variance 
(\ref{4}) can be rewritten as 
$\overline{[\tr\{\tilde\rho(t)\tilde A\} ]^2}$.
With the help of (\ref{1})  
and $\rho_{nm}:= \rho_{nm}(0)$ we finally obtain
\begin{equation}
\sa^2={\sum}' \tilde A_{jk} \rho_{kj} \tilde A_{mn} \rho_{nm} 
\, \overline{e^{i[E_j-E_k+E_m-E_n]t/\hbar}} \ ,
\label{5}
\end{equation}
where the sum ${\sum}'$ runs over all $j,k,m,n=0,1,2,...$ with 
$j\not = k$ and $m\not =n$.
Next we exploit the fact that $E_j-E_k+E_m-E_n$ 
vanishes for generic \cite{f9} Hamiltonians $H$ 
only for $j=n$ and $k=m$, given $j\not = k$ and $m\not =n$  
\cite{f3}.
Since the time averaged exponentials in (\ref{5}) 
vanish if $E_j-E_k+E_m-E_n\not = 0$ we can conclude that
\begin{eqnarray}
\sa^2={\sum}' |\tilde A_{mn}|^2 |\rho_{mn}|^2 \leq
\sum |\tilde A_{mn}|^2 |\rho_{mn}|^2  \ ,
\label{5b}
\end{eqnarray}
where the first sum runs over all $m \not=n$ and the second over
all $m,n$. 
With (\ref{5a}) and (\ref{3}) we thus obtain
\begin{eqnarray}
\sa^2\leq \sum \tilde A_{mn} \rho_{nn} \tilde A_{nm} \rho_{mm}  =
\sum\langle m|\tilde A\rho_{eq}|n\rangle \langle n|\tilde A\rho_{eq}|m\rangle  \ .
\nonumber
\end{eqnarray}
The sum over $n$ amounts to an identity operator and
that over $m$ yields $\tr\{[\tilde A\rho_{eq}]^2\}$.
This trace over the entire space $\hr$ can be
restricted to $\hr_+$ without changing its value,
as follows from  (\ref{1d}) and (\ref{3}).
Again, this trace remains unchanged if we now replace
$\tilde A$ by $\tilde A_p:=P\tilde AP$, where $P$ is the
projector onto $\hr_+$.
Next, we evaluate this trace with the help of the eigenvectors 
$|\chi_n\rangle$ of $\tilde A_p$ (restricted to $\hr_+$), yielding
\begin{eqnarray}
\sa^2\leq \sum \langle \chi_m|\rho_{eq}\tilde A_p|\chi_n\rangle 
\langle \chi_n|\rho_{eq}\tilde A_p|\chi_m\rangle \ .
\nonumber
\end{eqnarray}
Observing that 
$\tilde A_p|\chi_n\rangle = |\chi_n\rangle\langle\chi_n| \tilde A_p|\chi_n\rangle$ 
(since $|\chi_n\rangle$ is eigenvector of $\tilde A_p$)
and that 
$\langle\chi_n| \tilde A_p|\chi_n\rangle = \langle\chi_n| \tilde A|\chi_n\rangle$
(since $|\chi_n\rangle\in\hr_+$ and thus $P|\chi_n\rangle = |\chi_n\rangle$)
we can exploit (\ref{4a}) to obtain
\begin{eqnarray}
\sa^2\leq \da ^2 \sum \langle \chi_m|\rho_{eq}|\chi_n\rangle 
\langle \chi_n|\rho_{eq}|\chi_m\rangle \ .
\nonumber
\end{eqnarray}
The sum over $n$ yields the identity operator (on $\hr_+$) 
and that over $m$ amounts to $\tr\{\rho_{eq}^2\}$. 
With (\ref{1a}) and (\ref{3}) we finally arrive at
\begin{eqnarray}
\sa^2 \leq \da^2\, R^3 \ .
\label{8}
\end{eqnarray}

Considering $\tr\{\rho(t)A\}$ as a random variable with
uniformly distributed $t\geq 0$, a similar (but simpler) 
calculation as before yields for its mean value the
result $\overline{\tr\{\rho(t)A\}}=\tr\{\rho_{eq}A\}$.
Hence, (\ref{4}) is its variance and by combining
(\ref{8}) with Chebyshev's inequality 
\cite{wik}, we can conclude that 
\begin{eqnarray}
\mbox{Prob}\bigg(\big|\tr\{\rho (t)A \}-\tr\{\rho_{eq} A\}\big| \geq R \,\da  \bigg) 
\leq R \ .
\label{9}
\end{eqnarray}
In view of (\ref{1b}) it follows that for the overwhelming 
majority of times $t\geq 0$ the difference between 
$\tr\{\rho (t)A \}$ and $\tr\{\rho_{eq}A \}$ is way 
below the instrumental resolution limit $\dda$ for any 
experimentally realistic observable, 
see below (\ref{2}).
In other words, {\em the system looks exactly 
as if it were in the mixed state $\rho_{eq}$ for 
the overwhelming majority of times $t\geq 0$} \cite{f11},
though the ``true'' $\rho(t)$ is actually quite
different, see above (\ref{1c}).
This is our first main result.

Note that (\ref{9}) is still compatible with 
the recurrence property of $\tr\{\rho(t)A\}$
mentioned above (\ref{1c}) but implies 
that such excursions from  the
``apparent equilibrium state'' $\rho_{eq}$
must be exceedingly rare events.

Exactly the same ``apparent equilibration'' 
towards $\rho_{eq}$ emerges if one propagates 
$\rho(0)$ backward in time (keeping the 
system isolated).
Along the entire real $t$-axis, 
an initial condition $\rho(0)$ far from 
equilibrium thus closely resembles one of the
above mentioned rare excursions, just that
the location of this excursion is on purpose
chosen as the time-origin.
In other words, the quantum mechanical 
time inversion invariance is maintained,
but when starting out of equilibrium, 
an ``apparent time arrow'' emerges with
extremely high fidelity.

While (\ref{9}) provides a bound for 
the {\em relative} amount
of time the system exhibits notable
deviations from equilibrium, 
the typical duration of one given 
excursion, or equivalently, the characteristic
relaxation time of an out of equilibrium initial
condition $\rho(0)$ remains unspecified.
Since one can easily find examples
with arbitrarily large or small 
relaxation times,
any further quantification of the
relaxation process inevitably would require 
a considerably more detailed specification 
of the Hamiltonian $H$,
the initial state $\rho(0)$, and the
observable $A$.

Considering and estimating quantities like
(\ref{4}) is very natural and has a long 
tradition:
Merits and shortcomings of the early 
works are reviewed e.g. in \cite{ergod},
most notably Ludwig's approach \cite{lud58}.
In particular, many of them \cite{boc59,ergod}
involve an extra average over initial
conditions with the effect that any 
specific non-equilibrium seed 
must be excluded as ``potentially untypical'' 
from the general conclusions.
Turning to the more recent precursors,
Peres' approach \cite{per84} is
comparable to ours up to Eq. (\ref{5b}) but 
then proceeds with the conjecture 
that the $\tilde A_{mn}$ are pseudorandom 
matrix elements, statistically 
independent of the $\rho_{nm}$,
for which there are general arguments 
\cite{per84} and numerical evidence 
\cite{fei84} 
(and counter-evidence \cite{rig08})
but no proof.
For {\em pure} states, Srednicki obtained 
similar results \cite{sre96} by exploiting 
a common conjecture about the semiclassical 
behavior of classically smooth observables 
$A$ in systems with a fully chaotic classical 
limit.
Again, this conjecture is based on good 
arguments \cite{arg} but no proof.
Moreover, typical classical many-body systems 
are not expected to behave fully chaotic
\cite{pro94}.
Somewhat similar conclusion have been reached
even earlier by Deutsch \cite{deu91} via 
additional hand waving arguments.
Finally, rigorous results comparable
to (\ref{9}) are due to \cite{tas98,cra08},
but only for rather special Hamiltonians 
$H$ and initial conditions.

According to (\ref{3}) and the discussion
below (\ref{9}), expectation values become 
practically indistinguishable from
\begin{equation}
\tr\{\rho_{eq} A\} =\sum \rho_{nn}(0) A_{nn} 
\label{101}
\end{equation}
after initial transients have died out
(``equilibration'').
In how far is this in agreement with 
ESM, predicting \cite{lldiu} ``thermalization'', i.e.  
the appearance of the microcanonical ensemble 
$\rho^{\mic}$ instead of $\rho_{eq}$ in (\ref{101})?
In case $\rho^{\mic}$ and 
$\rho_{eq}$ yield measurable 
differences for experimentally 
realistic $\rho(0)$ and $A$, the 
``purely Quantum Mechanical'' prediction (\ref{101})
is commonly considered as ``more fundamental''
\cite{rig08,exp,rig07}.
In other words, {\em our derivation of ESM is 
complete}, provided the latter is valid itself.
In the opposite case, there is nothing to 
derive, but the quantity in (\ref{101}) 
still governs the (time-) typical behavior.
This the second main result of our Letter.

A first well known validity condition for ESM 
is a ``sharp energy $E$'', 
i.e. all the $\rho_{nn}(0)$ with $E_n \in I:=[E,\, E+\Delta E]$ 
sum up to almost unity, $\Delta E$ being small 
but still experimentally realistic \cite{lldiu}.
In particular, the $\rho_{nn}^{\mic}$ are constant 
for all $E_n\in I$ and zero otherwise \cite{lldiu},
and as a second (often tacit) validity condition for 
ESM, the resulting expectation values 
$\tr\{\rho^{\mic} A\}$ are assumed to be (practically) 
independent of the exact choice of $\Delta E$ and $E$.
Basically, this means that {\em the details of $\rho_{nn}(0)$ 
do not matter in (\ref{101})},
henceforth called property (P).
The same conclusion (P) follows from the 
equivalence of the microcanonical and canonical 
ensembles (for all energies $E$), considered 
as a self-consistency condition for ESM.
Clearly, property (P) is tantamount to replacing
$\rho_{eq}$ in (\ref{101}) by $\rho^{\mic}$.
Our first remark regarding (P) itself, is that 
no experimentalist can control the populations 
$\rho_{nn}(0)$ of the unimaginably dense 
energy levels $E_n$, apart from the very gross
fact that they are ``mainly concentrated within $I$''.
If the details would matter, not only ESM would 
break down but also reproducing measurements,
in particular in different labs, would 
be largely impossible. 
Second, one can readily construct 
observables and initial conditions, being
experimentally realistic according to our
definitions but still violating (P).
The fact that ESM is known to have an
extremely wide experimental applicability
implies that our so far notion of
``experimentally realistic'' is still
too general.
The simplest option seems to require/assume 
that the expectation values $A_{nn}=\langle n|A |n\rangle$
hardly vary within the energy interval $I$.
This is similar in spirit to classical 
coarse graining, and, in fact, 
is part of the already mentioned common 
conjecture about the semiclassical 
behavior of fully chaotic classical 
systems \cite{arg}.
Next, even if the $A_{nn}$ notably vary,
following Peres \cite{per84} the immense number 
of relevant summands in (\ref{101}) may
-- for ``typical'' $A$ and $\rho(0)$ -- 
lead to a kind of statistical 
averaging effect and thus a largely 
$\rho(0)$-independent final result.
All these conjectures about 
``truly realistic experimental conditions'' 
become even more compelling by considering that,
``canonically'', $A$ only affects a small 
sub-system, weakly coupled to a ``big'' rest, 
which can be readily traced out 
in (\ref{101}), with the effect of
an extra averaging step \cite{boc59,ergod,tas98,das03}.
Yet, the apparent universality of property (P)
and its relation to ``more basic'' system properties 
like ``ergodicity'' and ``(non-)integrability''
are still not very well understood 
\cite{rig08,ergod,per84,arg,wei92}.

Numerically, the validity and limits
of such conjectures and of ESM itself
have been exemplified e.g. in 
\cite{rig08,rig07,fei84,jen85}.
While the details are not yet settled, 
``equilibration'' in agreement with 
(\ref{101}) was seen in all cases.
Also the numerical observation that already quite small 
particle numbers often work surprisingly well 
is explained by (\ref{9}) in view of 
(\ref{1b}).

In the classical case, proving the
counterpart of the relation
$\overline{\tr\{\rho(t)A\}}=\tr\{\rho_{eq}A\}$ 
(see above Eq. (\ref{9})) is tantamount 
to the notorious ergodicity problem 
\cite{ergod}.
The next step, namely evaluating the classical
counterpart of (\ref{4}) remains as an even
more difficult open problem.


\begin{thebibliography}{7}
\bibitem{f9}
Though real systems are never strictly 
closed, our present standpoint (also
supported by numerical experience) 
implies that this should not be the 
pivotal point:
Otherwise, standard Quantum Mechanics 
alone would not be sufficient to model 
the (entangled) time evolution, and, on 
the other hand, ESM would not be applicable to 
strictly closed systems.
After including the most relevant 
perturbations ``from outside'' into 
the considered system,
it must be possible to theoretically
model it as strictly isolated from the 
rest of the world, with a well-defined
Hamiltonian $H$, possibly not known in detail 
but with generic properties.
\bibitem{f10}
Since $\rho (t)$ is non-negative and Hermitean,
$( \psi,\phi):=\langle\psi|\rho (t) +\epsilon|\phi\rangle$
is a well defined scalar product for any $\epsilon>0$.
Hence, Cauchy-Schwarz's inequality applies and by 
letting $\epsilon\to 0$ one obtains (\ref{5a}).
\bibitem{f11}
One readily sees that (\ref{3}) is indeed a well-defined
density operator and satisfies Liouville-von-Neumann's
equation.
\end{thebibliography}
\end{document}